\documentclass[11pt,twoside]{article}
\usepackage{./asp2010}

\resetcounters

\newcommand{\HI}{H\,{\sc i}}

\begin{document}

\title{Finding galaxies with unusual HI content}
\author{Helga~D\'{e}nes,$^{1,2}$ Virginia~A.~Kilborn,$^1$ and B{\"a}rbel~S.~Koribalski$^2$
\affil{$^1$Centre for Astrophysics \& Supercomputing, Swinburne University of Technology, P.O. Box 218, Hawthorne, VIC 3122, Australia}
\affil{$^2$Australia Telescope National Facility, CSIRO Astronomy and Space Science, P.O. Box 76, Epping, NSW 1710, Australia}}

\begin{abstract}
Observations show that galaxies in galaxy clusters are strongly influenced by their environment. There is growing evidence that some galaxies in groups show similar properties to galaxies in clusters, such as redder colours and gas deficiency, highlighting that environmental processes are also effective on galaxy group scales. The question is though, which mechanisms are important in low density environments? To answer this, we identify gas deficient galaxies to investigate recent or ongoing environmental processes, such as gas stripping. We are using scaling relations between the neutral hydrogen (\HI) content and optical properties of galaxies to identify galaxies with significantly less \HI\ than an average galaxy of the same type. We derive new, multi wavelength \HI-optical scaling relations using the \HI\ Parkes All Sky Survey (HIPASS) with optical and near infrared datasets. We also show our preliminary results from observations of a sample of 6 \HI-deficient galaxies, which we identified in low density environments. 

\end{abstract}

\section{Introduction}

The `morphology-density relation' \citep{Dressler1980} and the Butcher \& Oemler effect \citep{Butcher1987} suggest evolutionary processes that transform blue star forming galaxies into passive red early type galaxies in high density environments. It is also known that spiral galaxies in high density environments tend to have on average less \HI\ than galaxies of the same type and size in the field (e.g. \citealt{Davis1973, Giovanelli1985, Solanes2001}). This indicates that gas removal processes play an important role in the evolution of galaxies in high density environments. 

Gas stripping mechanisms are well studied in galaxy clusters, especially in the Virgo cluster (e.g. \citealt{Kenney2004, Vollmer2001, Chung2007}). Spiral galaxies in this environment are often \HI-deficient, i.e. they have at least 2 times less \HI\ mass than we would expect them to have based on their optical luminosity or diameter \citep{Haynes1984}. However, lower density environments, such as galaxy groups, are not nearly as well studied as galaxy clusters. Whilst there are a few studies showing that some late type galaxies in galaxy groups also appear to be \HI-deficient compared to field galaxies (e.g. \citealt{Kilborn2005, Sengupta2006, Rasmussen2006}), it is still not established if galaxy groups generally have \HI-deficient galaxies and which stripping mechanisms are dominant in low density environments. Galaxy-galaxy interactions are considered the main cause of gas stripping in group environments, since the density of the intra group medium (IGM) and the velocity dispersion of groups are thought to be insufficient for ram pressure stripping. Furthermore, only the most massive and evolved groups tend to have detected X-ray emission that would make thermal evaporation effective \citep{Zabludoff1998}. However, signs of possible ram pressure stripping were observed in galaxy groups (\citealt{Rasmussen2006, Westmeier2011}), which makes it clear that we still do not know enough about the efficiency of gas stripping mechanisms in low and intermediate density environments. 

\section{HI-optical scaling relations}
 
\begin{figure}[!b]
\centering
\includegraphics[width=100mm]{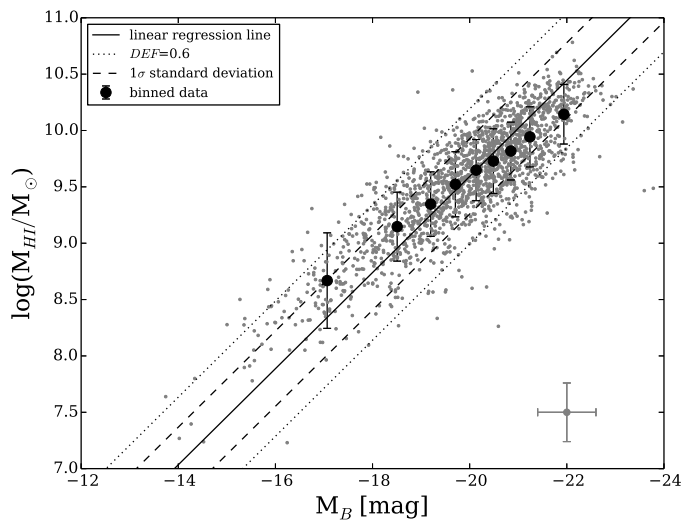}
\caption{The logarithm of \HI\ mass plotted against \textit{B}-band magnitude. The solid line is the fitted line to the data and the dashed line is the 1$\sigma$ standard deviation. The large symbols are magnitude bins with 200 galaxies each (except for the last bin). Galaxies below the dotted line are considered \HI-deficient and galaxies above it are considered to have \HI-excess.}
\label{fig:mag-logHI}
\end{figure}

\HI-optical scaling relations are a useful tool to investigate how different density environments influence the gas content of galaxies. We can characterise the \HI\ content of galaxies with scaling relations between the \HI\ content and other intrinsic properties of galaxies. Previous works investigating \HI-optical scaling relations for late type galaxies found that the optical diameter and luminosity correlate well with the \HI\ mass (e.g. \citealt{Haynes1984}; \citealt{Chamaraux1986}; \citealt{Solanes1996}). Initial scaling relations suffered from small sample sizes and optical selection biases. More recent studies (e.g. \citealt{Catinella2010}, \citealt{Toribio2011_2}) favour the use of \HI\ selected samples combined with optical properties from the Sloan Digital Sky Survey (SDSS). However \HI\ scaling relations are sensitive to the optical data. Scaling relations derived with SDSS data can only be reliably applied to galaxies in SDSS due to the unique photometric filters of SDSS. UV and SDSS photometry is not available for a large fraction of galaxies. In the view of upcoming large scale \HI\ surveys in the southern hemisphere this presents a problem, making it necessary to also establish scaling relations between other available optical data and the \HI\ content of galaxies. 

We derive \HI\ scaling relations for galaxies using the \HI\ Parkes All Sky Survey (HIPASS, \citealt{Barnes2001}) and a variety of optical and near-infrared luminosities and diameters. We use a multi-wavelength approach to determine scaling relations between the \HI\ content of galaxies and their diameter and luminosity, in 5 and 6 optical/IR wavebands respectively \citep{Denes}. The different bands are not uniformly affected by extinction and they probe different stellar populations of a galaxy. The large sky coverage of the catalogues from which we derive our scaling relations, makes them suitable to investigate how different environments influence the \HI\ content of galaxies. 

In Figure~\ref{fig:mag-logHI} we present one of our scaling relations, the relation between the logarithm of the \HI\ mass and the \textit{B}-band absolute magnitude. One of the applications of \HI\ scaling relations is to identify galaxies with anomalous \HI\ content  - i.e. \HI-deficient galaxies or galaxies with \HI-excess. We consider the \HI\ content of galaxies anomalous if they have at least 4 times more or 4 times less \HI\ than expected based on the scaling relations. These are the galaxies below and above the dotted lines in Figure~\ref{fig:mag-logHI}. These galaxies are excellent targets to investigate evolutionary processes in galaxies. For example \HI-excess galaxies may be accreting gas from their surrounding and \HI-deficient galaxies may be undergoing gas stripping processes.

\section{HI-deficient galaxies}

We selected a sample of the six most \HI\ deficient galaxies determined by the \textit{B}-band magnitude scaling relations from HIPASS for follow up observations to investigate the reason for their \HI-deficiency. These galaxies have at least 4 times less \HI\ than late-type galaxies with the same luminosity. This sample of \HI-deficient galaxies is particularly interesting since 4 of the sample galaxies (IC 1993, NGC 1473, NGC 1515, NGC 6808) are residing in lose galaxy groups one galaxy is in a galaxy triplet (ESO 009-G 010) and one galaxy is in the field (NGC 6699). \HI-deficient galaxies are well known and studied in dense galaxy clusters and in compact groups (e.g  \citealt{Giovanelli1985, Kenney2004, Vollmer2001, Verdes-Montenegro2001}), but only a handful of them is known in low density galaxy groups (\citealt{Chamaraux2004, Kilborn2005, Sengupta2006}). 

\begin{figure}[!ht]
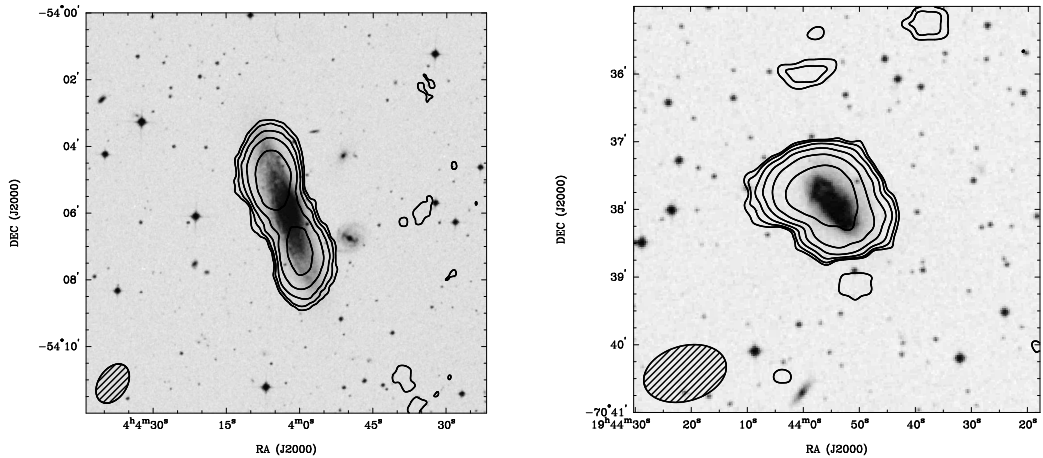

\begin{minipage}[b]{0.5\linewidth}
\center
\includegraphics[width=6cm, angle=270]{denesh_fig2a.eps}
\end{minipage}
\hspace{0.5cm}
\begin{minipage}[b]{0.5\linewidth}
\center
\includegraphics[width=6cm, angle=270]{denesh_fig2b.eps}
\end{minipage}
\caption{NGC 1515 (left) and NGC 6808 (right), ATCA HI contours are overlaid onto the optical B-band image from the Digitalised Sky Survey 2 (DSS2). Contour levels are 1, 3, 5, 7, 9 $\times 10^{19}$cm$^{-2}$}
\label{fig:HI-deficient-galaxies}
\end{figure}

We observed each galaxy for $\sim$10 hours with the Australia Telescope Compact Array (ATCA) 750 meter configuration. Our preliminary results show that one of the six galaxies, NGC 1515, appears to have a truncated and slightly warped \HI\ disk (Figure~\ref{fig:HI-deficient-galaxies}). Another of our sample galaxies, NGC 6808 apperas to have a slightly warped stellar disk, based on its Hubble Space Telescope (HST) image and a possible extension of its \HI\ disk pointing towards the centre of the galaxy group it belongs to (Figure~\ref{fig:HI-deficient-galaxies}). This suggest that the main gas stripping mechanism for this galaxy may be tidal interactions with group member galaxies. The other four galaxies are only partially resolved or unresolved in \HI. We conclude from these preliminary results that we need higher resolution \HI\ observations to identify the gas loss mechanisms responsible for the \HI-deficiency of these galaxies. We were awarded observing time on ATCA in this semester with a higher resolution configuration.

\acknowledgements We would like to thank the Astronomical Society of Australia for travel support. 

\bibliography{denesh-refs}

\begin{thebibliography}{}
\expandafter\ifx\csname natexlab\endcsname\relax\def\natexlab#1{#1}\fi
\expandafter\ifx\csname url\endcsname\relax
  \def\url#1{\texttt{#1}}\fi
\expandafter\ifx\csname urlprefix\endcsname\relax\def\urlprefix{URL }\fi
\providecommand{\eprint}[2][]{\url{#2}}

\bibitem[{{Barnes} et~al.(2001){Barnes}, {Staveley-Smith}, {de Blok}, \& {et
  al.}}]{Barnes2001}
{Barnes}, D.~G., {Staveley-Smith}, L., {de Blok}, W.~J.~G., \& {et al.} 2001,
  \mnras, 322, 486

\bibitem[{{Butcher} \& {Oemler}(1978)}]{Butcher1987}
{Butcher}, H., \& {Oemler}, A., Jr. 1978, \apj, 219, 18

\bibitem[{{Catinella} et~al.(2010){Catinella}, {Schiminovich}, {Kauffmann}, \&
  {et al.}}]{Catinella2010}
{Catinella}, B., {Schiminovich}, D., {Kauffmann}, G., \& {et al.} 2010, \mnras,
  403, 683. \eprint{0912.1610}

\bibitem[{{Chamaraux} et~al.(1986){Chamaraux}, {Balkowski}, \&
  {Fontanelli}}]{Chamaraux1986}
{Chamaraux}, P., {Balkowski}, C., \& {Fontanelli}, P. 1986, \aap, 165, 15

\bibitem[{{Chamaraux} \& {Masnou}(2004)}]{Chamaraux2004}
{Chamaraux}, P., \& {Masnou}, J.-L. 2004, \mnras, 347, 541

\bibitem[{{Chung} et~al.(2007){Chung}, {van Gorkom}, {Kenney}, \&
  {Vollmer}}]{Chung2007}
{Chung}, A., {van Gorkom}, J.~H., {Kenney}, J.~D.~P., \& {Vollmer}, B. 2007,
  \apjl, 659, L115

\bibitem[{{Davies} \& {Lewis}(1973)}]{Davis1973}
{Davies}, R.~D., \& {Lewis}, B.~M. 1973, \mnras, 165, 231

\bibitem[{{D\'{e}nes} et~al.(){D\'{e}nes}, {Kilborn}, \& {Koribalski}}]{Denes}
{D\'{e}nes}, H., {Kilborn}, V.~A., \& {Koribalski}, B.~A.(submitted)

\bibitem[{{Dressler}(1980)}]{Dressler1980}
{Dressler}, A. 1980, \apj, 236, 351

\bibitem[{{Giovanelli} \& {Haynes}(1985)}]{Giovanelli1985}
{Giovanelli}, R., \& {Haynes}, M.~P. 1985, \apj, 292, 404

\bibitem[{{Haynes} \& {Giovanelli}(1984)}]{Haynes1984}
{Haynes}, M.~P., \& {Giovanelli}, R. 1984, \aj, 89, 758

\bibitem[{{Kenney} et~al.(2004){Kenney}, {van Gorkom}, \&
  {Vollmer}}]{Kenney2004}
{Kenney}, J.~D.~P., {van Gorkom}, J.~H., \& {Vollmer}, B. 2004, \aj, 127, 3361

\bibitem[{{Kilborn} et~al.(2005){Kilborn}, {Koribalski}, {Forbes}, {Barnes}, \&
  {Musgrave}}]{Kilborn2005}
{Kilborn}, V.~A., {Koribalski}, B.~S., {Forbes}, D.~A., {Barnes}, D.~G., \&
  {Musgrave}, R.~C. 2005, \mnras, 356, 77

\bibitem[{{Rasmussen} et~al.(2006){Rasmussen}, {Ponman}, \&
  {Mulchaey}}]{Rasmussen2006}
{Rasmussen}, J., {Ponman}, T.~J., \& {Mulchaey}, J.~S. 2006, \mnras, 370, 453

\bibitem[{{Sengupta} \& {Balasubramanyam}(2006)}]{Sengupta2006}
{Sengupta}, C., \& {Balasubramanyam}, R. 2006, \mnras, 369, 360

\bibitem[{{Solanes} et~al.(1996){Solanes}, {Giovanelli}, \&
  {Haynes}}]{Solanes1996}
{Solanes}, J.~M., {Giovanelli}, R., \& {Haynes}, M.~P. 1996, \apj, 461, 609

\bibitem[{{Solanes} et~al.(2001){Solanes}, {Manrique},
  {Garc{\'{\i}}a-G{\'o}mez}, {Gonz{\'a}lez-Casado}, {Giovanelli}, \&
  {Haynes}}]{Solanes2001}
{Solanes}, J.~M., {Manrique}, A., {Garc{\'{\i}}a-G{\'o}mez}, C.,
  {Gonz{\'a}lez-Casado}, G., {Giovanelli}, R., \& {Haynes}, M.~P. 2001, \apj,
  548, 97

\bibitem[{{Toribio} et~al.(2011){Toribio}, {Solanes}, {Giovanelli}, {Haynes},
  \& {Martin}}]{Toribio2011_2}
{Toribio}, M.~C., {Solanes}, J.~M., {Giovanelli}, R., {Haynes}, M.~P., \&
  {Martin}, A.~M. 2011, \apj, 732, 93

\bibitem[{{Verdes-Montenegro} et~al.(2001){Verdes-Montenegro}, {Yun},
  {Williams}, {Huchtmeier}, {Del Olmo}, \& {Perea}}]{Verdes-Montenegro2001}
{Verdes-Montenegro}, L., {Yun}, M.~S., {Williams}, B.~A., {Huchtmeier}, W.~K.,
  {Del Olmo}, A., \& {Perea}, J. 2001, \aap, 377, 812

\bibitem[{{Vollmer} et~al.(2001){Vollmer}, {Cayatte}, {Balkowski}, \&
  {Duschl}}]{Vollmer2001}
{Vollmer}, B., {Cayatte}, V., {Balkowski}, C., \& {Duschl}, W.~J. 2001, \apj,
  561, 708

\bibitem[{{Westmeier} et~al.(2011){Westmeier}, {Braun}, \&
  {Koribalski}}]{Westmeier2011}
{Westmeier}, T., {Braun}, R., \& {Koribalski}, B.~S. 2011, \mnras, 410, 2217

\bibitem[{{Zabludoff} \& {Mulchaey}(1998)}]{Zabludoff1998}
{Zabludoff}, A.~I., \& {Mulchaey}, J.~S. 1998, \apj, 496, 39

\end{thebibliography}

\end{document}